\documentclass[preprint,showpacs,preprintnumbers,amsmath,amssymb]{revtex4-1}
\usepackage{graphicx}
\usepackage{dcolumn}
\usepackage{bm}

\begin{document}

\title{Synchronization of coupled nonidentical dynamical systems}

\author{Suman Acharyya}
\email{suman@prl.res.in}
\author{R. E. Amritkar}%
\email{amritkar@prl.res.in}
\affiliation{%
Physical Research Laboratory\\
Navrangpura, Ahmedabad-380009
}%

\date{\today}

\begin{abstract}
We analyze the stability of synchronized state for coupled nearly identical dynamical systems on networks by deriving an approximate Master Stability Function (MSF).
Using this MSF we treat the problem of designing a network having the best synchronizability properties. We find that the edges which connect nodes with a larger relative parameter mismatch are preferred and the nodes having values at one extreme of the parameter mismatch are preferred as hubs.
\end{abstract}

\maketitle

When two or more dynamical systems are coupled or driven by a common signal the systems may synchronize under suitable conditions\cite{sync_ref, PhysRevLett.64.821}. One can achieve different types of synchronization, such as complete synchronization \cite{PhysRevLett.64.821}, phase synchronization \cite{PhysRevLett.76.1804}, lag synchronization \cite{PhysRevLett.78.4193}, generalized synchronization \cite{PhysRevE.51.980} etc. Recently, there is considerable interest in the synchronization of coupled dynamical systems on a network \cite{PhysRep.366.1}. For coupled identical systems which give exact synchronization, Pecora and Carroll \cite{PhysRevLett.80.2109} have introduced
a master stability function (MSF) which can be calculated from a simple set of master stability equations and then applied to the study of stability of the synchronous state
of different networks. This general approach has become popular and has been used
in various studies of synchronization on networks \cite{PhysRevLett.89.054101,PRL.91.014101.03,PRE.72.057102.05,PRL.96.164102.06,PRE.81.026201.10}. Several works on different
networks have shown that small world and scale free networks show better synchronization properties \cite{PhysRevLett.89.054101,PRL.91.014101.03}.  

For coupled nonidentical systems, in general it is difficult to obtain exact synchronization. But, one can get synchronization of some generalized type \cite{PhysRevE.51.980}.
The parameter mismatch between different coupled systems can lead to desynchronization bursts and this is known as the bubbling transition \cite{PhysLettA.193.126,PhysRevE.54.1346}. 
Restrepo et el.
\cite{PhysRevE.69.066215} have studied the spatial patterns of such desynchronization bursts in networks. After the desynchronization burst the system returns to the synchronized state. Sun et al. \cite{epl.85.60011} determine the deviation from average trajectory as a function of the mismatch. However, for nonidentical systems, there is no general theory such as MSF, to study the stability of synchronization. 

In this paper we address the question of the stability of synchronization of coupled nearly identical systems on networks. By using the property of differential equations that the homogeneous part determines the exponential rates and treating the parameter mismatch in a first order perturbation theory, we derive master stability equation for coupled nearly identical systems. This master stability equation uses the homogeneous state and two parameters, $\alpha$ for
the network coupling and $\Delta$ for the mismatch in nonidentical systems. This allows us to define the MSF and study the stability properties of the synchronized state.

When one considers identical coupled systems the important question is about the type of network which gives better synchronization properties. When one considers coupled nearly identical systems, additional interesting and important questions arise. Which nodes are better chosen as hubs? Which edges give better synchronization? Using our MSF we find that for better synchronization nodes on one extreme of parameter mismatch are preferred as hubs and nodes with larger relative parameter mismatch are preferred for constructing edges.

Consider $N$ coupled dynamical systems,
\begin{eqnarray}
\dot{x}^{i} &=& f(x^{i},r_{i}) + \varepsilon \sum_{j=1}^{N}G_{ij}h(x^{j}); \qquad i=1,...,N \label{dyn-x-i}
\end{eqnarray}
where, $x^i(\in R^m)$ is an $m$ dimensional state vector of the system $i$, $f:R^m \rightarrow R^m$ gives the dynamics of an isolated system, $\varepsilon$ is a scalar coupling parameter and $h: R^m \rightarrow R^m$ is a coupling function, $G$ is the coupling matrix of the network, $r_i$ is some parameter which depends on the node $i$.

For the coupled identical systems, i.e. $r_i=r, \; \forall i$,
the synchronization manifold is defined by $x^1 = \cdots = x^N=x$ and is an invariant
manifold provided the
coupling matrix satisfies the condition that $\sum_j G_{ij} = 0, \; \forall i$.
With this condition, the synchronized state $x$, is a solution of the uncoupled dynamics, $\dot{x} = f(x)$.

The condition $\sum_j G_{ij} = 0$ ensures that $G$ has
one eigenvector
$e_1 = (1, \ldots, 1)^T$, with eigenvalue $\gamma_1=0$. This eigenvector
defines the synchronization
manifold. All the remaining eigenvectors
belong to the transverse manifold. The synchronized state is stable provided all the transverse Lyapunov exponents are negative.

Now, let us consider the case when the parameter $r_i$ depends on the node $i$.
Let the parameter mismatch be $\delta r_{i}=r_i - \tilde{r}$ where $\tilde{r}$ is some typical value of the parameters $r_i$.
In general, for nonidentical systems it is not possible to get an exact synchronization
of the type discussed above. Instead we get a generalized synchronization where there 
is a functional relationship between variables of the systems, e.g. $g(x^i, x^j) = 0$. The generalized synchronization is stable provided the largest transverse Lyapunov exponent is negative.

To determining the stability of this generalized synchronization, we do the linear stability analysis. In this analysis, we retain terms to second order in $z^i = x^i -x$ and $\delta r_i$. The reason for doing this will be clear shorty. Thus the dynamics of the deviation $z^i$ can be written as
\begin{eqnarray}
\dot{z}^i & = & D_x f(x,\tilde{r})z^i + \varepsilon\sum_{j=1}^{N}G_{ij}D_x h(x) z^j + D_{r} f(x,\tilde{r})\delta r_i 
\nonumber \\
& & + D_{r} D_x f(x,\tilde{r})z^i \delta r_{i} + \frac{1}{2}D_{r}^2 f(x,\tilde{r}) \delta r_i^2 + \ldots \label{mmsf03}
\end{eqnarray}
The terms corresponding to $(z^i)^2$ are not included since we will be interested in
the solution $z^i=0$ for finite $\delta r_i$.
As an equation for $z^i$, the RHS of Eq.~(\ref{mmsf03}) contains both homogeneous and inhomogeneous terms.
To a first approximation, the inhomogeneity won't affect the exponential rate of convergence of the trajectories to the synchronous solutions though it can shift the solution. 
To see this consider a general linear equation $Du = p(t)$, where $D$ is a differential operator. Let the solution be $u = u_h + g(t)$ where $u_h = \sum_i A_i h_i(t) exp(k_it)$ is the solution of the homogeneous equation $Du = 0$, and $A_i$ are constants. If $p(t)$ does not have any exponential dependence, then $g(t)$ cannot contain  
any additional exponential other than already in $u_h$, due to the property that the derivative of an exponential is also an exponential with the same exponent. For example, for $\dot{u} = -k u + p$, the solution of the homogeneous equation is $u_h(t)=u(0) e^{-k t}$ and of the inhomogeneous equation with constant $p$ is $u(t) = (u(0)-(p/k)) e^{-kt} + p/k$. We note that the inhomogeneity in the differential equation shifts the asymptotic solution but does not change the exponential.
In our case the stability of the synchronized state is governed by the largest transverse Lyapunov exponent, i.e. only by the exponential rates which are determined by the homogeneous equation. The inhomogeneous part will shift the solution.
In addition, while calculating the Lyapunov exponents, it is necessary that the shifted solution preserves the nature of the attractor so that the average expansion and contraction rates are not significantly affected by the shift. This can be assumed to be valid when different systems are in generalized synchrony since they are related to each other. This may also hold very near
the synchronization region but not far away from it.

Hence, to obtain the Lyapunov exponents, we consider the homogeneous equation obtained from Eq.~({\ref{mmsf03}).
\begin{equation}
\dot{z}^i = D_x f \ z^i + \varepsilon\sum_{j=1}^{N}G_{ij}D_x h \ z^j + D_r D_x f \ z^i \delta r_{i} \label{mmsf03b}
\end{equation}
This equation can be put in a matrix form as \cite{PhysLettA.296.204}
\begin{equation}
\dot{Z} = D_xf \ Z + \varepsilon D_{x} h \ Z \ G^T + D_r D_x f \ Z \ R \label{dyn-Z-r}
\end{equation}
where $Z =  (z^1, \ldots, z^N)$ is an $m \times N$ matrix and 
$R=\rm{diag}(\delta r_1,\ldots,\delta r_N)$ is an $N\times N$ 
diagonal matrix. 

Let $\gamma_k, \, e_k^R, \, k=1, \ldots, N$ be the eigenvalues and right eigenvectors of $G^T$.
Acting Eq.~(\ref{dyn-Z-r}) on $e_k^R$ and
using the $m$ dimensional vectors $\phi_k = Z e_k^R$,
we get
\begin{eqnarray}
\dot{\phi}_k & = & [D_xf + \varepsilon \gamma_k D_xu] \phi_k + D_r D_x f \ Z \ R \ e_k^R.
\label{dyn-phi-r}
\end{eqnarray}
In general, $e_k^R$ are not eigenvectors of $R$ and hence Eq.~(\ref{dyn-phi-r}) is not easy to treat. To solve
Eq.~(\ref{dyn-phi-r}) we use first order perturbation theory and write Eq.~(\ref{dyn-phi-r}) as
\begin{eqnarray}
\dot{\phi}_k & = & [D_xf + \varepsilon \gamma_k D_xu + \nu_k D_r D_x f] \phi_k
\label{dyn-phi-r-pert}
\end{eqnarray}
where $\nu_k = e_k^L R e_k^R$ is the first order correction and $e_k^L$ is the left eigenvector of $G^T$. 

Since both $\gamma_k$ and $\nu_k$ can be complex, treating them as complex parameters $\alpha = \varepsilon \gamma_k$ and $\Delta = \nu_k $ respectively,
we can construct the master stability equation as
\begin{equation}
\dot{\phi} = [D_xf + \alpha D_xh + \Delta D_rD_x f] \phi.
\label{msf-mod}
\end{equation}
For the coupled identical systems, the above equation reduces to the master stability equation given by Pecora and Carroll \cite{PhysRevLett.80.2109}.
We can determine
the MSF or $\lambda_{\rm max}$, which is the largest
Lyapunov exponent for Eq.~(\ref{msf-mod}), as a surface in the complex space defined by
$\alpha$ and $\Delta$ \cite{note_scope}.
The synchronized state is stable if the MSF is negative at each
of the eigenvalues $\gamma_k = \alpha$ and $\nu_k=\Delta$ ($k \neq 1$). This ensures that all the transverse 
Lyapunov exponents are negative.

We note that though the master stability equation (\ref{msf-mod}) uses the homogeneous state, it allows  
us to study the stability of the generalized synchronization in nonidentical systems.
The mismatch between the different systems is included through the parameter $\Delta$.

\begin{figure}
\begin{center}
\includegraphics[width = .9\columnwidth]{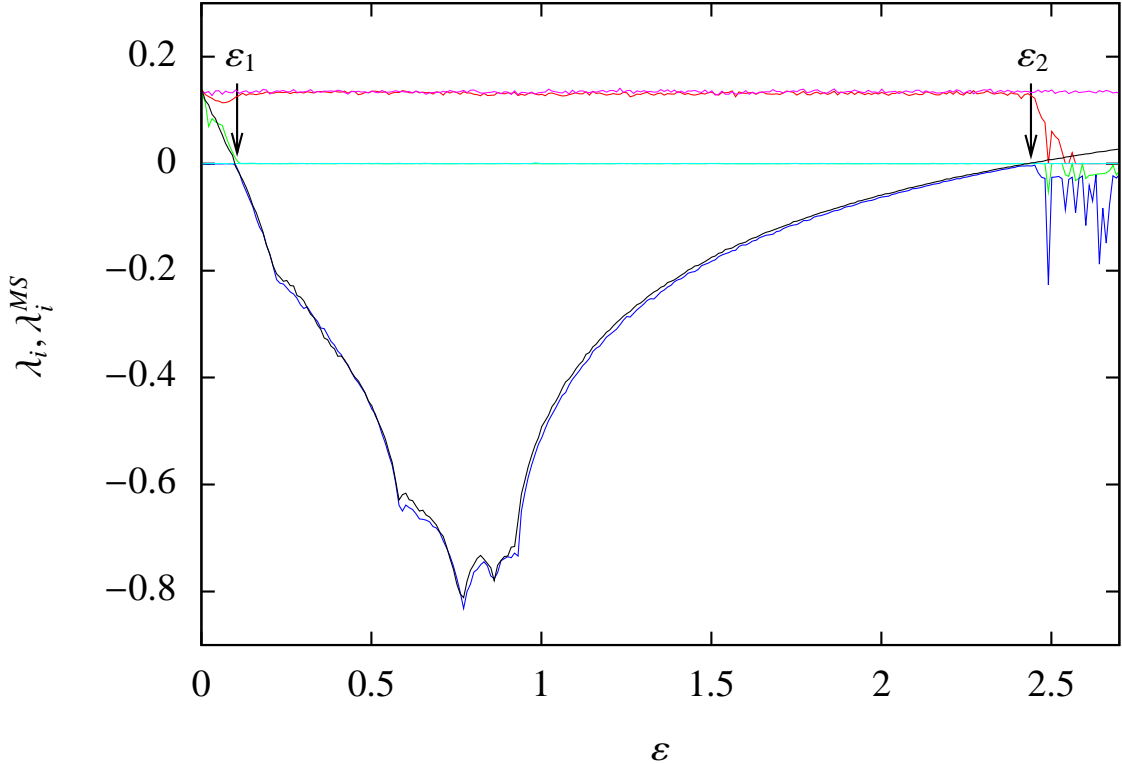}
\caption{\label{all_les} The figure shows the three largest Lyapunov exponents $\lambda_i, i=1,2,3$ (red, green and blue) and their estimated values $\lambda_i^{MS}$ obtained from the master stability equation (Eq.~(\ref{msf-mod})) (pink, cyan and black) as a function of $\varepsilon$ for two coupled R\"ossler systems with frequencies
$\omega_1 =1.05$ and $\omega_2 = 1.07$. Taking $\tilde{\omega} = 1.0$ we get $\Delta_1 = \Delta_2 = 0.06$ which are used in Eq.~(\ref{msf-mod}). R\"ossler parameters are $a_r=b_r=0.2, c_r=7.0$. The synchronous state is stable in the region given by $\alpha_1 < \alpha < \alpha_2$ indicated by the arrows.}
\end{center}
\end{figure}

To examine how well Eq.~(\ref{msf-mod}) allows the estimation of Lyapunov exponents, we calculate the Lyapunov exponents for the coupled
R\"ossler systems \cite{Rössler1976397} and compare them with those obtained from Eq.~(\ref{msf-mod}). Consider $N$ coupled chaotic R\"ossler systems with different frequencies,
\begin{eqnarray}
\dot{x}_{i} &=& -\omega_{i}y_{i}-z_{i} + \varepsilon\sum_{j=1}^{N} L_{ij}(x_{j}-x_{i}) \nonumber \\
\dot{y}_{i} &=& \omega_{i}x_{i} + a_{r}y_{i}  \label{ros0}\\
\dot{z}_{i} &=& b_{r} + z_{i}(x_{i}-c_{r}) \nonumber
\end{eqnarray}
where $\omega_i$ is the frequencies of the $i$-th oscillator and $L_{ij} = 1$ if the nodes $i$ and $j$ are coupled and zero otherwise and $L_{ii} = - \sum_{j \neq i} L_{ij}$.
For simplicity we restrict ourselves to symmetric coupling matrices $L$ so that the eigenvalues and hence $\alpha$ and $\Delta$ are real.

\begin{figure}
\begin{center}
\includegraphics[width = .9\columnwidth]{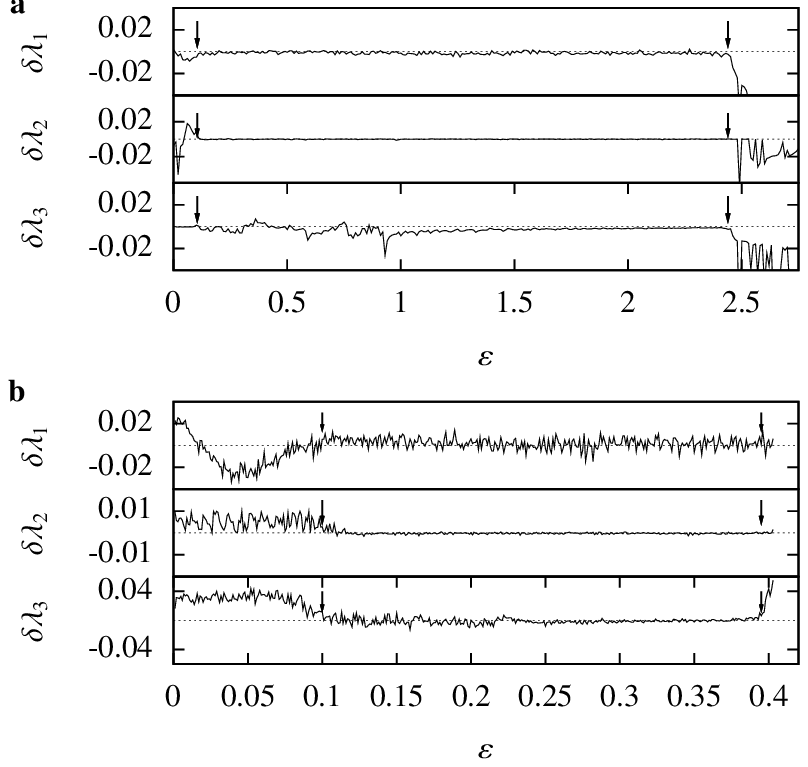}
\caption{\label{error_2} {\bf a}. The figure shows the difference $\delta \lambda_i = \lambda_i - \lambda_i^{MS}$ for the three largest Lyapunov exponents as a function of the coupling constant $\varepsilon$ for two coupled R\"ossler systems with parameters as in Fig~1. {\bf b}. The figure shows the difference $\delta \lambda_i$ for the three largest Lyapunov exponents as a function of $\varepsilon$ for sixteen randomly coupled R\"ossler systems having different internal frequencies $\omega_i$. We find that the differences are small in the synchronization region.}
\end{center}
\end{figure}

We first consider two coupled R\"ossler oscillators. Fig.~1
plots the three largest Lyapunov exponents, $\lambda_i, i=1,2,3$, as a function of the coupling strength $\varepsilon$ and their estimated values $\lambda_{i}^{MS}$ from Eq.~(\ref{msf-mod}). Fig.~2a plots the difference $\delta \lambda_i = \lambda_i - \lambda_i^{MS}$ as a function of $\varepsilon$ for these Lyapunov exponents. The region when the third largest Lyapunov exponent $\lambda_3<0$, corresponds to the synchronization region and in this region it is the largest transverse Lyapunov exponent. 
From Figs.~2a, we find that the differences $\delta \lambda_i$ are small in the synchronization region and very close to it. Though only three exponents are plotted in the figure, the differences are small for the other Lyapunov exponents. Fig.~2b plots the difference $\delta \lambda_i$ as a function of $\varepsilon$ for the three largest Lyapunov exponents for a random network of sixteen nodes. Again we observe that the errors are small
in the synchronization region. Thus, we find that the master stability equation (\ref{msf-mod}) can estimate the actual Lyapunov exponents for the synchronized state reasonably well. 

\begin{figure}
\begin{center}
\includegraphics[width = 0.9\columnwidth]{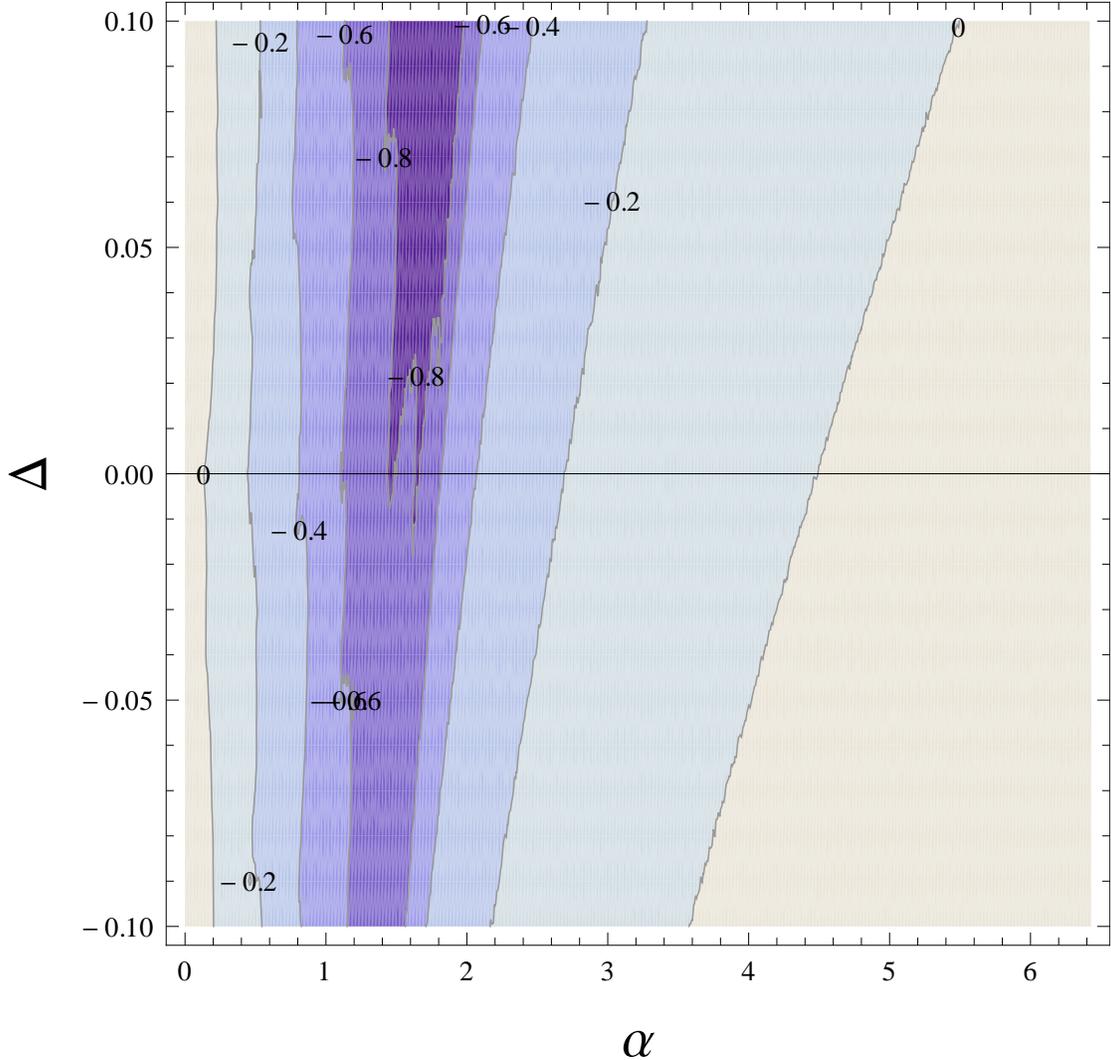}
\caption{\label{mmsf_omega} The master stability function $\lambda_{max}$ for R\"{o}ssler system is plotted as a contour plot in the parameter plane $(\alpha,\Delta)$. The stability region is given by the ``V'' shape region bordered by the $0$ contours from both sides.}
\end{center}
\end{figure}

Now, we consider the MSF, $\lambda_{max}$, which is the largest
transverse Lyapunov exponent. It can be calculated using Eq.~(\ref{msf-mod}). In Fig.~\ref{mmsf_omega} we plot $\lambda_{max}$ in the parameter plane $(\alpha,\Delta)$ as a contour plot for R\"ossler system. From the figure we can see that the stability region increases with the parameter $\Delta$.

We now demonstrate the utility of the master stability function by considering the
problem of construction of an optimized network which gives best synchronization properties. To construct the optimized network we adapt Monte Carlo optimization method \cite{metropolis:1087} and rewire the edges of the network to construct a network that shows best synchronizability, i.e. the largest interval $l_{\varepsilon}$ of the coupling constant
$\varepsilon$ which shows synchronization. 

We start with a system of nearly identical coupled R\"ossler oscillators as in Eq.~(\ref{ros0}) on a connected network of $N$ nodes and $E$ randomly chosen edges. In each Monte Carlo step we rewire one edge. If the rewired network increases the stability interval $l_{\varepsilon}$ of the synchronized state, then it is chosen with probability one, otherwise it is accepted with probability $e^{\beta  (l_{\varepsilon}^{new} -l_{\varepsilon}^{old})}$ where
$\beta$ is the inverse temperature.

We now investigate two questions. In the optimized network, which edges are more preferable and which nodes have larger
number of connection or act as hubs?

To investigate the question of which nodes act as hubs, we define the correlation coefficient between the frequency and the degree of a node as $\rho_{\omega k} = \frac{<(k_{i}-<k_i>)(\omega_{i}-<\omega_i>)>}{\sqrt{<(k_{i}-<k_i>)^2><(\omega_{i}-<\omega_i>)^2>}}$
where $k_i = -L_{ii}$ is the degree of node $i$. Fig.~\ref{correlation}{\bf a} shows $\rho_{\omega k}$ (solid line) as a function of Monte Carlo steps. For the random network $\rho_{\omega k} = 0$. We find that $\rho_{\omega k}$ increases and saturates to a positive value. Thus, in the synchronized optimized network the nodes which have larger frequencies have more connections and are preferred as hubs. The reason for this is the ``V'' shape of the stability region in
Fig.~\ref{mmsf_omega}, i.e. the stability range increases as $\Delta$ increases. We have also
investigated a case were an opposite behavior is obtained. If instead of the frequency,
we make the parameter $a_r$ in Eq.~(\ref{ros0}) node dependent, then the stability region in the plot of MSF similar to Fig.~\ref{mmsf_omega}, has an inverted
``V'' shape. In this case in the optimized network, nodes which have smaller values of
$a_r$ have more connections and are preferred as hubs.

To investigate the question of which edges are preferred, we define the correlation coefficient between the absolute frequency differences between two nodes and the edges as $\rho_{\omega a}= \frac{<(A_{ij}-<A_{ij}>)(|\omega_{i}-\omega_j|-<|\omega_{i}-\omega_j|>)>}{\sqrt{<(A_{ij}-<A_{ij}>)^2><(|\omega_{i}-\omega_j|-<|\omega_{i}-\omega_j|>)^2>}}$ where $A_{ij}=1$ if nodes $i$ and $j$ are connected and 0 otherwise.
Fig.~\ref{correlation}{\bf b} shows $\rho_{\omega a}$ as a function of Monte Carlo steps.  We find that $\rho_{\omega a}$ increases from 0 (the value for the random network) and saturates. Thus, in the synchronized optimized network the pair of nodes which have a larger
relative frequency mismatch are preferred as edges for the optimized network. 
Again, the reason for this preference of edges is probably 
the conical shape of the stability region in Fig. 3. The
edges are to be chosen so that the parameter $\Delta$ increases and the stability region increases.

\begin{figure}[t]
\includegraphics[width = .9\columnwidth]{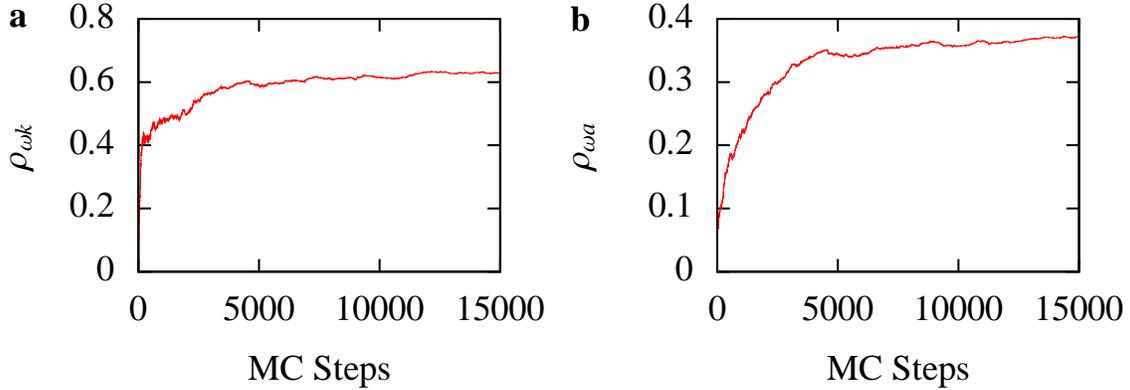}
\caption{\label{correlation} The figure plots the correlation coefficient $\rho_{\omega k}$ (figure a) and $\rho_{\omega a}$ (figure b) as a function of the Monte Carlo steps of optimization for
32 coupled R\"ossler systems. We see that both $\rho_{\omega k}$ and $\rho_{\omega a}$ increase and saturate to positive values.}
\end{figure}

To conclude we have developed the Master Stability Function (MSF) approach for coupled
nonidentical systems. We use the property of differential equations that the homogeneous part is mainly responsible for the exponential dependence of the variables. The parameter mismatch is treated in a first order perturbation theory. Our MSF uses the homogeneous state but it still allows us to study the stability properties of generalized  synchronization for nonidentical systems. Using MSF, we construct
optimized networks with better synchronization properties by rewiring the network keeping the number of edges constant. We find that in the optimized network the nodes having parameter mismatch at one extreme depending on the shape of stability region in MSF plot, have more
edges and are preferred as hubs and the pair of nodes which have a larger relative parameter
mismatch are preferred for constructing edges.


\begin{thebibliography}{99}
\bibitem{note_scope} The scope and limitations of the master stability approach are discussed in
Ref.~\cite{PhysRevLett.80.2109} and the same apply in our case also.
\end{thebibliography}
\end{document}